\documentclass[twocolumn,showpacs,preprintnumbers,amsmath,amssymb]{revtex4-2}
\usepackage{chemformula} % Formula subscripts using \ch{}
\usepackage{graphicx}
\usepackage[colorlinks=true,linkcolor=blue,citecolor=blue]{hyperref}
\usepackage{color}
\usepackage{braket}
\usepackage{physics}

\begin{document}

\title{Femtosecond laser induced creation of G and W-centers in silicon-on-insulator substrates}

\author{Hugo Quard$^1$}\email{hugo.quard@insa-lyon.fr}
\author{Mario Khoury$^2$}
\author{Andong Wang$^3$}
\author{Tobias Herzig$^4$}
\author{Jan Meijer$^4$}
\author{Sebastian Pezzagna$^4$}
\author{Sébastien Cueff$^1$}
\author{David Grojo$^3$}
\author{Marco Abbarchi$^{2,5}$}\email{marco.abbarchi@im2np.fr}
\author{Hai Son Nguyen$^{1,6}$}
\author{Nicolas Chauvin$^1$}
\author{Thomas Wood$^1$}

\affiliation{$^1$Univ Lyon, Ecole Centrale de Lyon, CNRS, INSA Lyon, Universit\'e Claude Bernard Lyon 1, CPE Lyon, CNRS, INL, UMR5270, 69130 Ecully, France} 
\affiliation{$^2$Aix Marseille Univ, CNRS, Universit\'e  de Toulon, IM2NP, UMR 7334, F-13397 Marseille, France}\affiliation{$^3$Aix-Marseille University, CNRS, LP3, UMR7341, 13009 Marseille, France}
\affiliation{$^4$Division of Applied Quantum Systems, Felix-Bloch Institute for Solid-State Physics, University Leipzig, Linn\'eestrasse 5, 04103 Leipzig, Germany}
\affiliation{$^5$Solnil, 95 Rue de la R\'epublique, 13002 Marseille, France}
\affiliation{$^6$Institut Universitaire de France (IUF)}

\date{\today}

\begin{abstract}
The creation of fluorescent defects in silicon is a key stepping stone towards assuring the integration perspectives of quantum photonic devices into existing technologies. Here we demonstrate the creation, by femtosecond laser annealing, of W and G-centers in commercial silicon on insulator (SOI) previously implanted with $^{12}$C$^+$ ions. Their quality is comparable to that found for the same emitters obtained with conventional implant processes; as quantified by the photoluminescence radiative lifetime, the broadening of their zero-phonon line (ZPL) and the evolution of these quantities with temperature. In addition to this, we show that both defects can be created without carbon implantation and that we can erase the G-centers by annealing while enhancing the W-centers' emission. These demonstrations are relevant to the deterministic and operando generation of quantum emitters in silicon.

\end{abstract}


\maketitle

\section{Introduction}

Point defects in silicon have been intensively studied over the past few years for the creation of silicon-based quantum devices\cite{khoury_bright_2022}. Thanks to the device-friendly environment, single photon sources or spin-photon interfaces could be readily integrated within existing electronic and photonic devices. Such components are key devices for quantum computing, quantum networks or for the implementation of quantum cryptography protocols.

\begin{figure*}[ht!]
    \centering
    \includegraphics[width=0.9\linewidth]{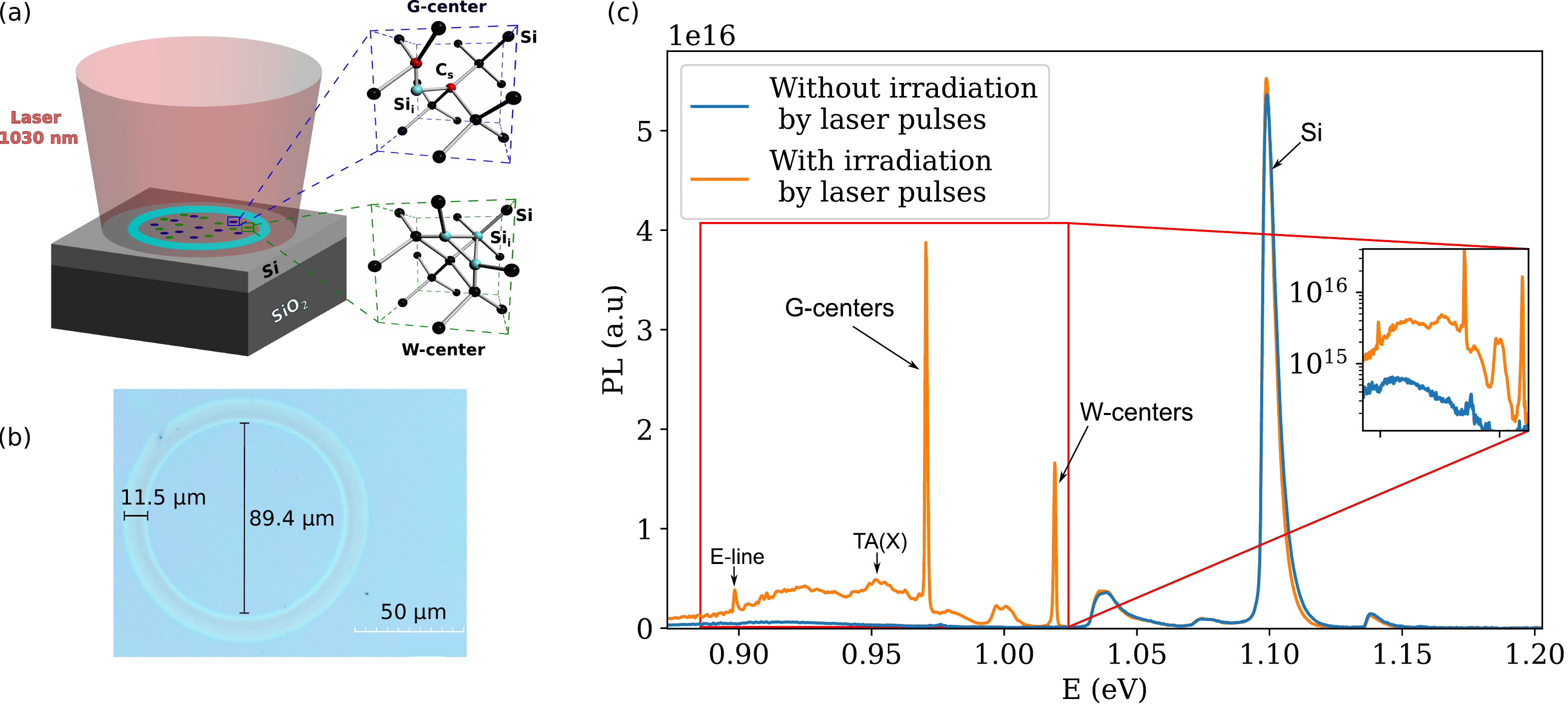}
    \caption{\textbf{(a)} Schematic representation of the laser irradiation process used to create G and W-centers.\textbf{(b)}Optical microscope image of an area irradiated with 3 pulses having an energy of 218 $\mu$J. \textbf{(c)}Comparison of the PL spectra at 12K of an area not irradiated by laser pulses and an area irradiated by 3 pulses.}
    \label{Figure1}
\end{figure*}

Among all the different optically-active defects in silicon \cite{skolnick_defect_1981,Thonke1981,safonov_interstitial-carbon_1996,bergeron_silicon-integrated_2020,durand_broad_2021,redjem_defect_2022,hollenbach_wafer-scale_2022,ivanov_effect_2022,redjem_single_2020},  two often studied are W \cite{davies_1018_1987,Buckley2017,Davies1987,carvalho_density-functional_2005,buckley_optimization_2020,Menold2021} and G-centers  \cite{song_bistable_1990,murata_high-density_2011,berhanuddin_co-implantation_2012,wang_g-centers_2014,berhanuddin_generation_2015,Beaufils2018,timerkaeva_structural_2018,beaufils_exaltation_2019,redjem_spectroscopie_2019,zhu_manipulating_2020,hollenbach_engineering_2020,udvarhelyi_identification_2021,hollenbach_photonic_2021,khoury_light_2022,prabhu_individually_2022,baron_single_2022,khoury_bright_2022,lefaucher_cavity-enhanced_2023,komza_indistinguishable_2022,Zhiyenbayev23}. W-centers are composed of three interstitial Si atoms. Recent theoretical works revealed that these defects probably have a $I_3$-V configuration \cite{Baron2022} and the optically active form of a G-center is made of two C atoms in substitutional sites linked to an interstitial Si atom \cite{song_bistable_1990} as represented in Fig \ref{Figure1}a).

Numerous procedures can be used to fabricate W or G-centers such as silicon \cite{Buckley2017} or carbon implantation followed by a proton irradiation \cite{Berhanuddin2012,Beaufils2018}, electron irradiation \cite{Thonke1981}, Focused Ion Beam (FIB) implantation \cite{hollenbach_wafer-scale_2022}, pulsed ion beams \cite{Schenkel2022},  and reactive ion etching \cite{Weber1991,Buyanova1995}. All these methods involve the use of ion implanters or accelerators. They are relatively bulky and expensive approaches that require several steps (e.g. for G-centers: carbon implant, annealing, and proton implant) in order to create a light emitter. Furthermore, the most advanced methods using focused ion beams\cite{hollenbach_wafer-scale_2022} are intrinsically stochastic and the number of quantum emitters per ion impact is not precisely controlled.   

A further method, that can be used to create point defects, is laser annealing also called laser doping. The use of a laser has been shown to be a viable approach for integrating quantum emitters \cite{Murata2010} in $\delta$-doped Si \cite{Murata2010} or in p-doped Si \cite{Monflier2021}. Both W and G-centers can be obtained in p-doped Si \cite{Menold2021} as well as in n-type Si implanted with $^{29}$Si$^+$ ions \cite{skolnick_defect_1981}. However, the creation of these defects was usually unintentional, induced with laser pulses longer than 30 ns and in Si substrates.

In this paper we address the creation of W and G-centers by femtosecond laser annealing, with the process being investigated for the first time on SOI substrates, taking both C-doped and pristine SOI wafers as starting points. The quality of these defects is confirmed by CW and time resolved photoluminescence measurements. The temperature dependence of the emitters, their broadening and lifetimes are  comparable to those reported with standard fabrication methods, accounting for the high quality of our approach based on fs laser pulses. We also demonstrate that the C-implant step is not necessary to create the light emitters. Finally, by low temperature annealing, we can selectively erase the G centers while improving the quality of the W centers' emission.

\section{Sample description and experimental setup}
The samples are SOI wafers featuring a 220 nm top silicon layer and a 2 µm buried oxide layer, some of which were implanted with $^{12}$C$^+$ ions. The beam energy was set to 34 keV in order to implant the carbon ions halfway into the top silicon layer and two different doses have been explored, namely $1\times 10^{12}$ and $1\times 10^{13}$ ions/cm$^2$. The implantation was followed by a flash annealing under N$_2$ atmosphere during 20 s at 1000°C to remove lattice damage \cite{Berhanuddin2012}.

Then the samples were irradiated by laser pulses focused with a 750mm plano-convex lens under ambient conditions, differing from location to location by the number of pulses used (from 1 to 5) and by the energy of the pulses (95,143,175 and 218  $\mu$J), to supply the energy necessary to reorganise the crystal lattice and create emitting defects in silicon. This step is schematically represented in Fig \ref{Figure1}a). The Gaussian laser beam used was centered at 1030 nm with pulses of a duration inferior to 200 fs. Its waist was $w_0=178$ $\mu$m and the repetition frequency of the pulses was 1kHz.  The samples have been created in pairs, sharing the same carbon implantation parameters, of which only one of the two underwent a second annealing under N$_2$ atmosphere during 5 min at 125 $^{o}$C after the implantation.

PL measurements were performed at low temperature, the samples being cooled down to 12~K with a closed-cycle liquid-helium cryostat. The optical pumping was performed with a continuous wave laser diode at 405~nm focused onto the sample with a spot diameter of $\approx$75~$\mu$m. A Cassegrain objective was used to collect the PL emission with a numerical aperture between 0.15 and 0.4. The collected signal was focused onto an optical fiber connected to a spectrometer coupled with a liquid nitrogen-cooled InGaAs detector enabling spectral detection from 900 to 1600~nm. For time-resolved PL, a 515~nm laser emitting 200~fs pulses and a 54~MHz repetition rate was used for the optical pumping and the detection was performed with an InGaAs photodiode.

\section{Results}
\subsection{Creation of G and W-centers}
For the highest pulse energy investigated (218 $\mu$J), the laser pulses created a ring visible under an optical microscope at the surface of the irradiated parts of the sample around the point of impact of the laser beam, as shown in Fig.\ref{Figure1}b). The creation of this circle is specific to ultrafast quenching conditions accessible with femtosecond laser irradiation. It results from the melting of the top layer of Si which resolidifies in different states due to the temperature gradient which spatially varies the solidification kinetics. The ring corresponds to amorphous Si whereas as the central part is constituted of recrystallized Si \cite{bonse_modifying_2004,Monflier2021}. The outer contour of the ring shown in Fig1 (a) corresponds to a local fluence of about 330 mJ/cm$^2$ that matches well with the measured threshold for Si surface amorphization in a previous work using the same laser \cite{Garcia2021}.

The PL signal from these areas reveals the creation of both G and W-centers, as shown by the orange curve in Fig.\ref{Figure1}c), exhibiting the zero phonon lines (ZPLs) of the two emitters, at 1.019~eV for W-centers and 0.97~eV for G-centers, as well as their typical phonon sidebands\cite{Beaufils2018,Thonke1981,baron_detection_2021}. 

On the same sample we also collected the PL emission in areas that were not targeted by the laser (blue curve in the Fig.\ref{Figure1} c)). The corresponding spectrum exhibits only the typical Si signal at 1.1~eV accompanied with the relative phonon sidebands.

For lower laser fluences, the ring is not observed (Supplementary Information \cite{supp}) along with a lack of PL emission from W- and G-centers, suggesting the crucial role of the melting and recrystallization steps in rearranging the atoms to form the emitters.

We also studied the influence of the number of pulses on the intensity, the position and the width of the ZPLs (Supplementary Information \cite{supp}). No trend can be seen from one sample to another, which suggests that at each pulse there is melting of the silicon and therefore the destruction of pre-existing defects. New defects are then created during recrystallization.

\subsection{Influence of the carbon dose}

\begin{figure}[!ht]
\centering
\includegraphics[width=0.9\linewidth]{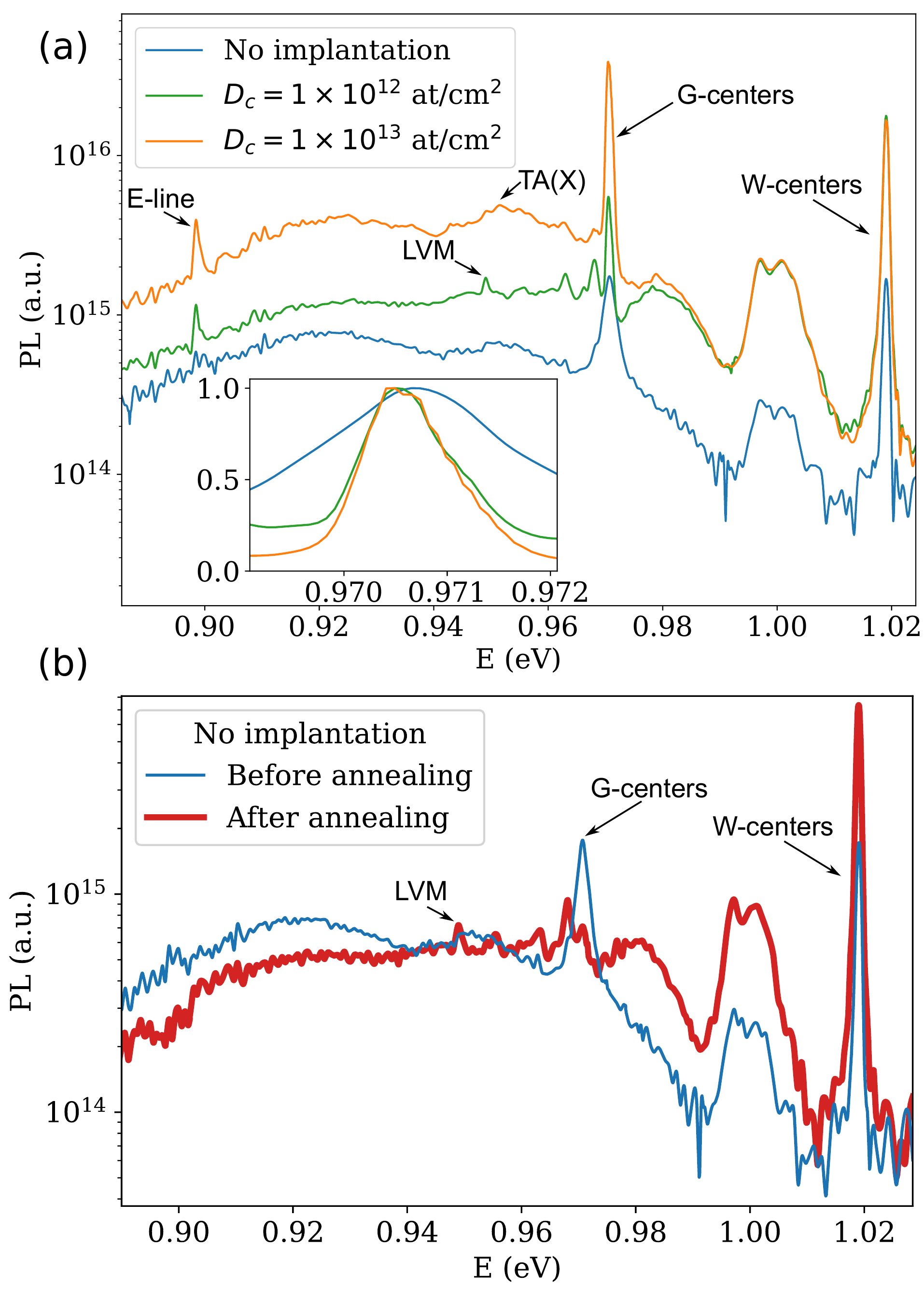}
  \caption{All the spectra were measured at 12 K with excitation by a laser diode at 405 nm. \textbf{(a)} Macro-PL spectra obtained for samples irradiated by 3 laser pulses with an energy of 218~$\mu$J. Each sample is differentiated by the dose of implanted carbon $D_C$. The inset represents the normalized ZPL of G-centers. \textbf{(b)} Comparison between macro-PL spectra obtained for two samples irradiated by 3 laser pulses with an energy of 218~$\mu$J . The two samples have not undergone carbon implantation and one of them underwent a second annealing under N$_2$ atmosphere for 5~min at 125$^{o}$C .}
  \label{Figure2}
\end{figure}

In this section we focus on the influence of the dose of implanted carbon on the PL of the centers. Fig. \ref{Figure2}a) presents the PL spectra of three samples which differ by the carbon dose used during the implantation. For each spectrum the Si signal is of the same order of magnitude which allows a direct comparison between spectra. 

The ZPL intensity of the G-centers increases when raising the carbon dose, which is consistent with previous findings\cite{Berhanuddin2012} for emitters created after proton irradiation. 

For pristine samples (ie. no carbon implantation), the signals of the G- and W-centers appear with an amplitude that is about an order of magnitude lower with respect to their implanted counterparts. We interpret the creation of G-centers in carbon-free samples as an effect of incorporation of residual C present in small quantities in the upper layers of Si (e.g. deposited during the manufacturing process of the SOI wafer). We verified that, by oxidation of the top Si layer in a rapid thermal processor, the signal from G-centers disappears, confirming that the C contamination comes from the sample surface (not shown).

As for the W-centers, this result shows that the implantation step disrupts the crystalline organisation and create more interstitial Si. However, even if defects are created in smaller numbers in the pristine sample, they have the same optical properties as those obtained with the implantation step; namely ZPL with the same full width at half maximum (FWHM) (see Supplementary Information \cite{supp}). This is not the case for G-centers for which the ZPL of the non implanted sample is broader and slightly blue-shifted (see Fig. \ref{Figure2}a)).

\subsection{Effect of annealing}
We now investigate the effect of a flash annealing on the PL signal of these laser-created emitting centers.  Non-implanted samples after flash annealing display only W-center emission that is enhanced by a factor of four with respect to non-annealed samples whereas that of G-centers disappears (Fig. \ref{Figure2} b)). 

The annihilation of G-centers after a flash annealing is also observed for carbon implanted samples and the resulting spectrum is comparable to that one obtained for pristine samples when normalised by the maximum of the ZPL of W-centers. Therefore, we demonstrate that we can create and isolate W-centers without any implantation step with optical properties comparable to those obtained with implantation.

\subsection{Recombination dynamics of G-centers}

\begin{figure}[ht!]
\centering
  \includegraphics[width=0.8\linewidth]{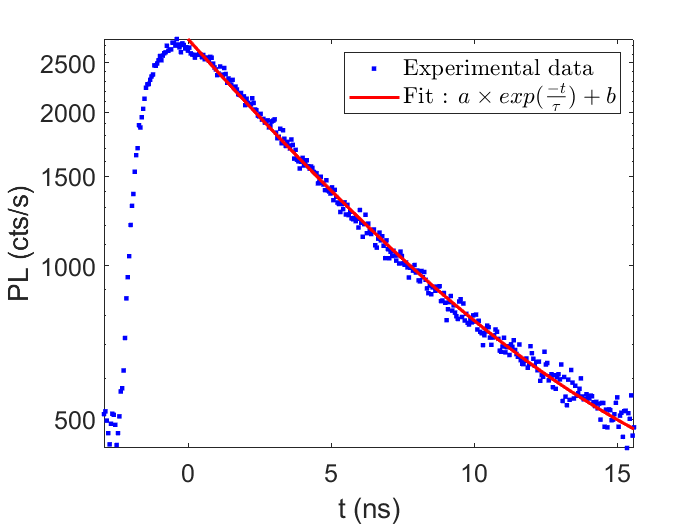}
  \caption{Time-resolved PL signal obtained at 12~K with a pulsed laser at 515~nm with a shortpass filter in energy (0.99 eV) to eliminate  the contribution of the ZPL of W-centers and a large part of their phonon sideband. The experimental data (blue points) are fitted by a mono-exponential function adjusted by a constant $b$ (red curve).} 
 \label{Figure3}
\end{figure}

We performed time-resolved PL measurements on the brighter sample, namely the sample with the highest dose of C implanted, filtering the PL signal with a short-pass filter at 0.99 eV to eliminate the major contribution of W-centers (Fig.\ref{Figure3}). The PL decay is well fitted by a mono-exponential function (plus a constant) providing a characteristic lifetime of 5.9~ns, consistent with conventional G-centers obtained by co-implant of carbon and proton irradiation for which the lifetime is about 5-6 ns \cite{Beaufils2018,lefaucher_cavity-enhanced_2023}. Therefore, G-centers created with both protocols have the same excited state lifetime. Note that the radiative yield of G-centers is less than 10\% at 30K \cite{lefaucher_cavity-enhanced_2023}. Therefore, the measured decay rate is strongly related to non-radiative channels even at cryogenic temperatures.

The constant used to adjust the fit is ascribed to the contribution of a longer decay time corresponding to the phonon-sideband of the W-centers: these defects have lifetimes ranging from 3 ns to 30~ns \cite{Baron2022}. Owing to our  excitation pulses being repeated every 18 ns, an overlap of the PL decays coming from the phonon-sideband of W-centers can contribute to the time resolved spectra. We cannot measure the lifetime of W-centers because of the overlap issue previously described.

\begin{figure*}[ht!]
    \centering
    \includegraphics[width=0.9\linewidth]{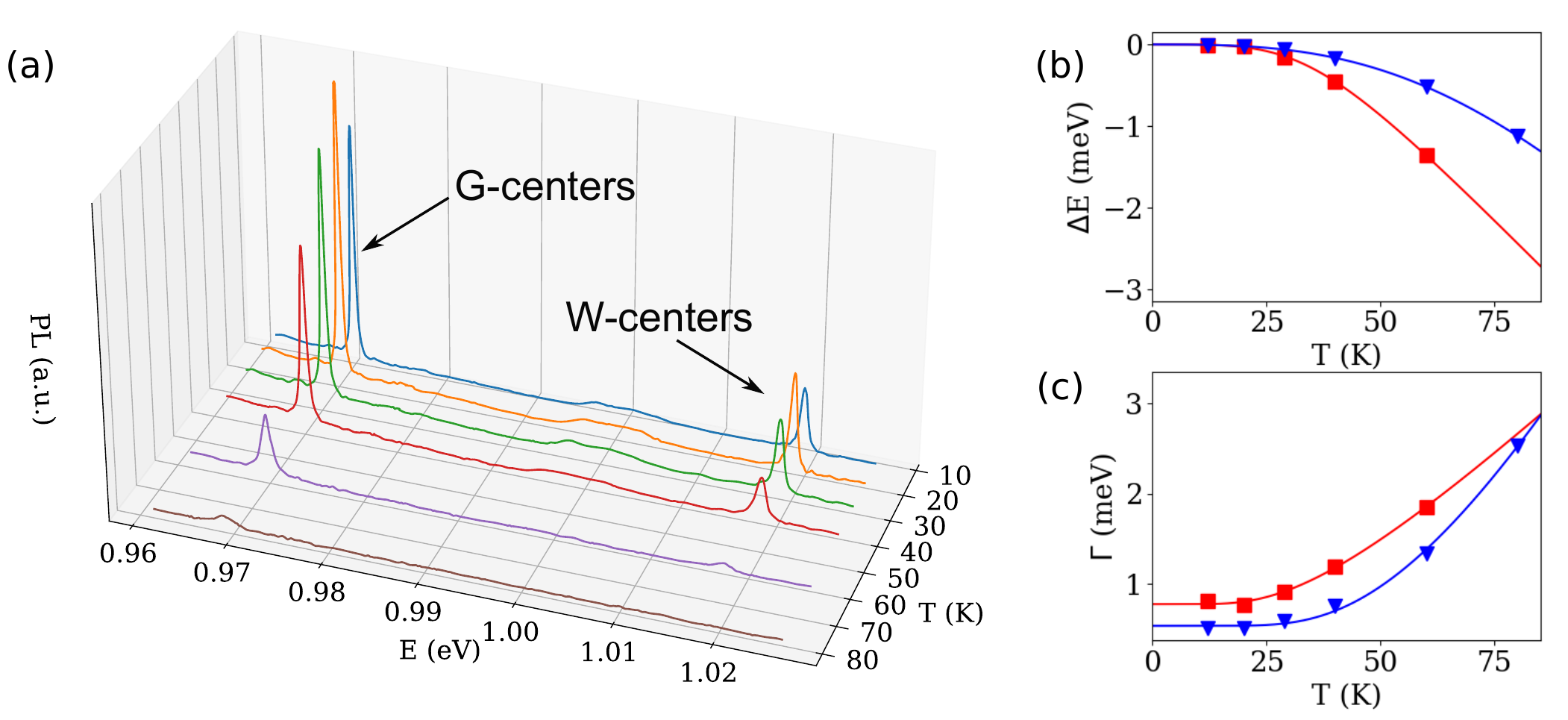}
    \caption{\textbf{(a)} Temperature-dependant PL spectra centred around the ZPLs of G and W-centers in the range of temperature from 12 K to 80 K. \textbf{(b)} Variation of the energies at the center of the ZPLs of G-centers (blue triangles) and W-centers (red squares) as a function of temperature. The symbols represent the experimental data and the lines the fits given by the equation \eqref{eq : delta E}. The fitting parameters are $E_{0W} = 1018.69\pm0.02$ meV, $\alpha_W=0.06\pm0.01$ meV/K, $\Theta_W=76\pm 15$ K, $P_W=4.2\pm1.1$ for W-centers and $E_{0G} = 970.194\pm0.003$ meV, $\alpha_G=0.07\pm0.01$ meV/K, $\Theta_G=186\pm 29$ K, $P_W=3.0\pm0.1$ for G-centers. \textbf{(c)} Evolution of the FWHM of the ZPLs of G-centers (blue triangles) and W-centers (red squares) as a function of temperature. The symbols represent the experimental data and the lines the fits given by the equation \eqref{eq : gamma}.  The fitting parameters are $\Gamma_{0W} = 0.78\pm0.03$ meV, $a_W=5\pm1$ meV, $\Omega_W=9\pm 1$ meV for W-centers and $\Gamma_{0G} = 0.54\pm0.03$ meV, $a_W=20\pm5$ meV, $\Omega_W=17\pm 1$ meV for G-centers.}
    \label{Figure4}
\end{figure*}

\subsection{Temperature dependence of the ZPL}

The ZPL energy as a function of temperature is represented in Fig.\ref{Figure4} b). The experimental data are fitted with the expression proposed by \citeauthor{passler_basic_1997} \cite{passler_basic_1997}:
\begin{eqnarray}\label{eq : delta E}
    \Delta E(T) &=& E_{ZPL}(T)-E_0 \nonumber\\
    &=&-\frac{\alpha\Theta_p}{2}\left[\sqrt[p]{1+\left(\frac{2T}{\Theta_p}\right)^p}
-1\right]
\end{eqnarray}
where $E_0$ (meV) is the limit of the ZPL energy when $T\to 0$ K, $\alpha$ (meV/K) is the slope of the curve, namely the entropy, for $T\to\infty$, $\Theta_p$ (K) is the average temperature of the phonons and p is a dimensionless parameter. 

The thermal redshift for W-centers evolves proportionally to $T^4$ \cite{cardona_isotope_2005} whereas for G-centers to $T^3$, consistent with previous observations \cite{Beaufils2018}. These fits provide the average temperature of the phonons coupled with the defects. If we convert the obtained values to energies we find E$_{phW}=7\pm 1$ meV for W-centers and E$_{phG}=16\pm 2$ meV for G-centers. For the latter, the obtained energy is close to that of TA phonons at the X point of the Brillouin zone \cite{ivanov_effect_2022}, namely 20 meV. Therefore we conclude that the defect coupled preferentially with TA(X) phonons. As for W-centers, the obtained value does not correspond to a typical phonon energy in silicon and it is lower than the TA(X) energy. A possible explanation for this observation will be given in the following paragraphs.

Fig.\ref{Figure4} c) represents the FWHM of the ZPLs as a function of temperature. The experimental data are well fitted by the following model that describes the broadening of a ZPL with heating \cite{rudin_temperature-dependent_1990}:
\begin{equation}\label{eq : gamma}
    \Gamma=\Gamma_0+a\left[\exp(\frac{-\Omega}{k_B T})-1\right]^{-1}
\end{equation}
where $\Gamma_0$ is the zero-temperature limit of the FWHM and the second term accounts for the coupling between phonons and the emitters, $a$ represents the intensity and $\Omega$ the typical energy of this coupling. 

The zero-temperature limit obtained for G-centers with our procedure, namely $\Gamma_{0G}=0.54\pm 0.03$ meV, is of the same order of magnitude as that obtained for an ensemble of G-centers created by proton irradiation ($0.3\pm0.03$ meV) \cite{beaufils_optical_2018}. The slightly larger value obtained in our case can be explained by the simultaneous presence of G-centers and Si self-interstitials which could slightly broaden the signal obtained for the ensemble of defects. 

For ensembles of W-centers, to date, no study has been conducted for ZPL broadening. The study of individual W-centers \cite{baron_single_2022} showed a zero-temperature limit below 0.1 meV. However, the energy differences between the ZPLs from one defect to another is within 1 meV owing to local variations in the  emitter environment. Therefore, the value of $\Gamma_{0G}=0.78\pm 0.03$ meV we obtained is consistent with single defect investigation when taking into account the ZPLs' dispersion in energy.

%THIS PARAGRAPH IS NOT CLEAR: 
For G and W-centers, the typical energies $E_{ph}$ and $\Omega$ we obtained with the fits of the broadening and the redshift of ZPLs overlap, taking into account uncertainties. This allow us to give a probable explanation for the low value obtained for $E_{phW}$. Indeed, the typical energy obtained with the fit of the broadening ($\Omega_W$) could also be interpreted as the activation energy of the excitonic transition to the first excitonic state from the ground-state \cite{borri_exciton_2003}, which could explain why we obtain a value lower than the typical energy of phonons in Si.

\section{Discussion}

The relevance of our work lies in the novel possibility to deterministically create quantum emitters in Si with near unity yield : in analogy with SiC and diamonds, fs laser pulses can be, in principle, used to form the emitters \textit{in situ} and \textit{operando}, while monitoring the emission from newly formed defects on a pulse-by-pulse basis \cite{chen2019,chen2019b}. This method also allows to reduce the area of creation of the emitters down to the size the laser spot as shown in Fig. \ref{Figure1} which demonstrates that the emitting centers are created only in the irradiated areas. Even if we do not reach the precision of FIB, this method is cheaper and easier to implement which is promising for the large scale creation of emitters. Moreover we demonstrate that W and G-centers created in C-implanted Si have optical properties and recombination dynamics in adequacy with the literature, which proves that we obtained emitters with the same quality as those obtained by others fabrication methods. 

We also demonstrate that the implantation step is not necessary to create W and G-centers, as we collect the PL signature of both centers from  pristine samples. Such samples contain only residual carbon incorporated in the Si during the manufacturing of the wafer, which implies a low concentration of carbon. This leads to the creation of low-density G-centers, which is confirmed by the low intensity of the ZPL of G-centers but also by its broadening and blue-shift compared to the implanted samples (see inset of Fig. \ref{Figure2}). Indeed \citeauthor{Zhiyenbayev23} demonstrates that the increase of internal strain due to an high dose of implanted carbons leads to a redshift the ZPL of G-centers and reduces their inhomogeneous broadening \cite{Zhiyenbayev23}. As for W-centers, they have the same optical properties in implanted and non-implanted samples. We demonstrate a way to annihilate G-centers while slightly enhance the PL of W-centers by carrying out an annealing at 125$^{o}$C during 5 min. This phenomenon is opposite to what was expected, since it has been demonstrated that the intensity of the ZPL of G-centers created with carbon implantation followed by proton irradiation can be increased by a factor of 8 with such thermal treatment \cite{berhanuddin_generation_2015}. It is worth noting that in the sample of \citeauthor{berhanuddin_co-implantation_2012} only G-centers were created whereas in our case we have both W and G-centers which could explain the difference in behavior of G-centers.

\section{Conclusion}
In this paper, we demonstrate for the first time the creation of G and W-centers simultaneously in carbon-implanted SOI by femtosecond laser annealing. The quality of these defects are comparable to those obtained by the usual, well established methods in the literature. We also demonstrated that we can create G and W-centers without any implantation step. However, G-centers created with this method have a ZPL broader than those obtained with carbon implantation. Furthermore, we demonstrated that an annealing at low temperature annihilates G-centers while slightly enhancing the PL emission of the W-centers. Therefore, we proved that we can create and purify W-centers of good quality without any implantation steps, which allows us to create W-centers at low cost, on-demand and in a restricted area of a size close to the cross section of the laser spot. This represents a step forward for the deterministic creation of W-centers in photonic structures. Indeed, with this method it is possible to precisely position the defects in the structures and it is also conceivable to have a control on the density of defects created by studying in more detail the influence of the number of pulses and their energy. This last aspect could even lead to the study of single emitters, assuming that we manage to create the W centers in sufficiently low density.\\

\textit{Acknowledgement:}
This research was funded by the EU H2020 FET-OPEN project
NARCISO (No. 828890), the French National Research Agency
(ANR) through the projects ULYSSES (No. ANR-15-CE24-0027-01), OCTOPUS (No. ANR-18-CE47-0013-01), European Research Council (ERC) under the European Union’s Horizon 2020 research and innovation program (Grant Agreement No. 724480). The authors thank the Nanotecmat platform of the IM2NP institute. 

\bibliography{biblio}

\end{document}